**Predictors of the Sense of Presence in an Immersive Audio Storytelling Experience,**

**a Mixed Methods Study**

<u>**Non-peer-reviewed pre-print version**</u>

Isabelle Verhulst[1], Rich Hemming[1], Adam Ganz[1], James Bennett[1], Rachel Donnelly[2],

Dawn Watling[1], and Polly Dalton[1].

[1]Royal Holloway University of London, UK

[2]Imperial War Museums, UK

**Author Notes**

Isabelle Verhulst 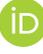 https://orcid.org/0000-0003-4603-0435

Rich Hemming 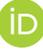 https://orcid.org/0000-0002-8521-1725

Adam Ganz 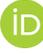 https://orcid.org/0000-0002-2167-689X

Dawn Watling 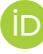 https://orcid.org/0000-0003-3727-4198

Polly Dalton 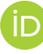 http://orcid.org/0000-0002-1999-5306

We have no conflicts of interest to disclose. Correspondence concerning this article should be addressed to Isabelle Verhulst, Royal Holloway University of London, Department of Psychology, Egham Hill, Egham, TW20 0EX, United Kingdom. Email: Isabelle.verhulst@rhul.ac.uk.

# Overview

**Abstract**

This study examined which variables predicted the sense of presence ('being there') in an immersive audio experience, with a focus on the impacts of immersion technology (headphones with spatialised sound versus speaker with 2D stereo sound), the nature of the audio experience, the narrative content, participants' emotional and cognitive engagement and personal characteristics such as age and gender. Museum visitors listened to a story on the 'One Story, Many Voices' immersive audio installation. A convergent mixed-methods design was used, including multiple regression analysis of survey data (n = 185) and relational analysis of interview data (n=9). This study found mixed methods support from both surveys and interviews to suggest that presence was predicted by the audio quality (especially the audio being perceived as different from other audio stories they had listened to in the past), cognitive engagement (especially being totally absorbed), the narrative (the story and how it was told) and some support was found for emotional engagement (especially feeling a connection with the storyteller). Effects of the environment and immersion technology on presence received only either quantitative or qualitative support. No influences of personal characteristics were found. Findings are relevant for academics, immersive experience (XR) commissioners and developers.

*Keywords: Presence, Predictor, Audio, XR, Immersive, Storytelling*

## 1. Presence in an Immersive Sound Experience

**1.1. What Predicts Presence in Immersive Experiences?**

Immersive technologies (XR, for example, Virtual Reality (VR) or spatialised sound experiences) are used ever more frequently in heritage organisations (Jung, et al., 2016; Mohd Noor Shah & Ghazali, 2018). Immersive experiences can bring stories and objects of the past to life, create more emotional, social, and authentic experiences, and stimulate learning (Kidd & McAvoy, 2019; Lee et al., 2020). Within such experiences, presence, or the sense of 'being there', has been identified as a key concept, defined as "… a state of consciousness, the (psychological) sense of being in the virtual environment." (Slater & Wilbur, 1997, p. 4.) For example, presence can have positive effects on users' enjoyment and behavioural intent such as whether to (repeat) visit (Bennett et al., 2021; Suh & Prophet, 2018; Sylaiou et al., 2010; Tussyadiah et al., 2018). Therefore, understanding the determinants of presence is important in designing strong user experiences. However, very little research has addressed presence predictors in immersive audio experiences, which is the aim of this study.

Many factors can influence or predict presence (for reviews see for example, Cummings & Bailenson, 2016; Felton & Jackson, 2022, Oh et al., 2018, on social presence; van den Berg (thesis), 2020, and Souza et al., 2021 on virtual reality presence). The following overview of earlier work focuses on factors that are relevant to the immersive audio experience that formed the basis of this study.

First, the extent of the immersion that is afforded by an experience, in terms of the technical aspects of the device used in the experience, or "how well it can afford people real-world sensorimotor contingencies for perception and action" (Slater & Sanchez-Vives, 2016, p. 37) is known to positively affect presence in general (for example, Cumming & Bailenson, 2016, Ma et al., 2023, Oh et al., 2019); although findings are mixed about the effect of immersion on social presence (Oh et al., 2019).In relation to audio experiences in particular, immersion may vary between different audio delivery mechanisms, for example headphones versus speakers, and this may influence presence. Pettey et al. (2008) argued this may be due to the social distance of a soundscape: headphones may be more immersive because they create a more 'intimate' (less public) soundscape experience for the listener. This is in line with findings from Kallinen and Ravaja (2007) that participants liked listening to headphones more than to speakers when listening to the news, even though participants reported that the quality of the sound was the same, and from Bracken et al. (2010) who found that headphones created higher presence levels than loudspeakers when hearing a video clip.

Second, the nature of the audio experience provided (whether there is any audio at all, and if so, its quality) is likely to affect presence. Research suggests that adding audio to a silent experience generally increases presence (e.g. Cumming & Bailenson, 2016), as does adding bass (Freeman & Lessiter, 2001; Lessiter & Freeman, 2001). However, relatively little is known about the effect of immersive sound on presence (for example, spatialised sound).

Of the few studies to date, most have found positive effects of immersive sound on presence in virtual environments (VEs; as reviewed, for example, by Bosman et al., 2023). For example, Poeschl et al. (2013) created a virtual visual 3D scene, in which participants saw a forest clearing from a first-person perspective, either with spatialised sound or without sound. The soundscape (used only in the sound condition) included ten virtual sound sources, some of which were moving (for example, the sound of a person walking past the participant) whereas others were fixed (for example, the sound of a waterfall). Participants reported significantly higher presence in the spatialised sound (vs. no sound) condition. Adding spatialised audio has also been shown to improve presence in a walking VR experience. Kern & Ellermeier (2020) found that including either a spatialised ambient soundscape or self-triggered footsteps could improve overall presence and realism while walking in a VR experience. However, as all these studies contrasted spatialised sound against a no sound condition, no conclusions can be drawn about the influence of the spatialisation itself from these studies.

Surround sound was compared to stereo sound in a video game study by Skalski and Whitbred (2010). They compared self-reported presence levels of 74 participants playing a first-person shooter game, with either multi-channel surround sound (Dolby 5.1) or two channel sound (Dolby stereo). Participants filled out the Temple Presence Inventory (TPI, Lombard & Ditton, 2009) which measures five presence elements: engagement, spatial presence, social richness, perceptual realism, and social realism. They found higher scores for surround sound than stereo for engagement, spatial presence, social richness, and perceived realism, but not for social realism, which the authors suggest may have been due to the violent content or small sample size.

A broader range of audio types have been compared in other studies, generally finding higher presence with higher levels of audio immersion. For example, mono (1-channel), stereo (2-channels), Dolby surround (multiple-channels), and 3D audio (realistic audio representation) were used to present wasp sounds to participants in a study by Brinkman et al. (2015). In their first test (audio only, without visuals, not in VR), presence increased as audio immersion increased. However, they reported one surprising result: presence was lower for Dolby surround headphones than stereo, which they suggest may be due to the Dolby algorithm simulating the surround sound. In their second test (audio and visuals, in VR), they found presence was higher in either stereo or 3D sound than without sound, but with no significant difference between stereo and 3D audio (mono and Dolby surround were not included in this test). However, these findings must be interpreted with caution given the relatively small sample sizes n=22 and 25). Hendrix and Barfield (1995) compared no sound, non-spatialised, and spatialised audio conditions in a virtual environment consisting of a room with common items, viewed on a projection screen, with stereoscopic view and headtracking. They found that adding spatialised sound (using HRTF) to a display with no sound significantly increased presence (but not perceived realism; experiment one), and that presence (not realism) was also significantly higher when using spatialised sound than when using non-spatialised sound (experiment two). Binaural audio, a technique used in both audio capture and playback that intends to create a 3D audio experience for the listener, can also create higher presence levels than monoaural audio (Västfjäll, 2003; Hendrix & Barfield, 1995).

There may be a ceiling to the effect of spatial sound fidelity on presence. For example, adding sound (speakers, headphones, and headphones with subwoofer) to an online video game increased users' presence in a study by Scorgie and Sanders (2002), with headphones having a similar effect on presence as a 5.1 surround sound speaker system, but including a subwoofer did not increase presence further. Similarly, Riecke et al. (2009) added different types of sound to a large rotating image and found that, compared to no sound, adding non-moving (mono or ambient) sound had no effect, adding sound that moved in sync with the visual increased presence, but increasing the spatial sound fidelity from a five-channel home theatre standard to a high standard with 5-degree resolution did not further increase presence.

The perceived realism of sounds is a complicated concept in audio and thought to be at least as important as that of visuals for presence, if not more so (Magnenat-Thalmann et al., 2005). Indeed, realism is seen by some researchers as part of presence (for example, in the Igroup Presence Questionnaire, IPQ, Schubert et al., 2001), although others see it as a separate construct (for example, Hendrix & Barfield, 1995). Certain types of audio content can improve realism and/or presence more generally. For example, in a qualitative study by Rogers et al. (2018) with 12 VR games participants, ambient sounds tended to be discussed as linked to realism and presence, whereas background music was often linked with emotional responses, and narration and sound effects were mentioned less often. By contrast, Jørgensen (2017) found that background music in video games affected both players' presence and perceived realism. In a study on audio fidelity, Lindquist et al. (2020) found that both ambient sound (generic background sounds, relevant to the scene) and realistic (detailed) sound (stereo recordings from relevant areas enriched with sounds from a database) increased perceived realism (versus no sound), and realistic (detailed) sound increased realism more so than ambient sound.

Whereas the research discussed so far has assessed the impacts of immersive audio within visually depicted environments, positive effects of spatialised sound on presence have also been shown in audio-only experiences. For example, presence and emotional realism were higher when listening to music in an auditory environment with six audio channels, than with mono or stereo sound (Västfjäll, 2003). Spatialised sounds of people seeming to approach a participant in a sound booth created higher presence than non-spatialised sound, both when using a presence questionnaire and when tested with physiological measurements (Kobayashi et al., 2015).

Although most studies have found positive effects of immersive sound on presence, some studies presented mixed or null results. For example, mixed results were found in Lessiter and Freeman's (2001) study. In experiment one, participants gave higher presence ratings when using a 5.1 surround sound version than mono or stereo sound; but in experiment two, with two to five audio channels, they did not find a significant difference in presence, possibly due to the potential ceiling effect mentioned earlier. A few researchers did not find any effect of immersive sound on presence. For example, Larsen and Pilgaard (2015) did not find a difference in presence between stereo and 3D spatialised video game sounds, and Narciso et al. (2019) did not find a difference when testing stereo and spatialised sound in an immersive 360 video user experience. In Rogers et al.'s (2018) second (quantitative) study, they tested four audio versions that decreased in audio dimensionality from the full audio version (with narration, sound effects, ambient noises, and music) down to a version with voice and sound effects only. They did not find a significant difference between these conditions on their measure of 'immersion' (the Immersion Experience Questionnaire, IEQ, Jennett et al., 2008), which has the following factors: challenge, cognitive involvement, control, emotional involvement, and real−world dissociation. This is unexpected as well-established research suggests that sensory richness can positively affect presence in VEs (Rogers et al., 2018). It is not clear if this null-result is due to their relatively small between-subjects sample size (n=40), if motion-sickness played a role (reported by 16 of 40 participants), and/or if there is little gain due to audio quality after a certain point, like Dinh et al. (1999) suggested for vision. Overall, the above suggests that the presence (vs. absence) of audio as part of an experience, as well as the level of spatialisation of any audio presented, can significantly affect people's sense of presence. When compared to other immersive technology features that can affect presence, Cumming and Bailenson (2016), in their review of 115 effect sizes from 83 studies, found that sound (the absence or presence and/or the quality of sound) had a larger effect size than image quality, but smaller than update rate, user-tracking level, field of view of visual displays, and stereoscopic visuals. However, most of this work focuses on audiovisual experiences, with very little research examining the effects of sound spatialisation within audio-only environments. This therefore forms a key focus of the current study.

Based on the existing research, we also expected a range of other factors to predict presence, namely the experience content (narrative, e.g. Felton & Jackson, 2022; Gorini et al., 2011; Riches et al., 2019), emotional engagement (e.g., Felton & Jackson, 2022, Pallavicini et al., 2020), Riches et al., 2019, Riva et al., 2007), cognitive engagement (e.g., Schubert & Crusius, 2002, Schubert et al., 2001, Schaik et al., 2004), participants' personal characteristics (e.g., Felton & Jackson, 2022, Oh et al., 2018) and specifics of the environment (e.g., Slater & Steed, 2000, Liebold et al., 2017).

Much of the abovementioned knowledge comes from lab-studies using VR which, by enclosing the eyes, partially isolates the user from the real world. This helps the user to suspend their disbelief that they are in a virtual world, rather than where their real bodies are, which in turn helps to maintain the sense of presence (Slater & Usoh, 1993). In audio experiences, participants may have access to conflicting real-world visuals throughout the experience, or may choose to limit their visual input by, for example, closing their eyes. Given the importance of visuals in determining presence in VR, the absence of any matching visuals in audio experiences (and indeed the likelihood of conflicting visuals) means that presence is likely to be experienced differently in audio experiences and is thus likely to be influenced by different factors. In this study, we investigated if these factors predicted participants' presence in a real-life audio installation setting. 1.2. Research Approach

The current study uses a mixed-methods approach, in a real-life setting, adding to the understanding of how presence can be created in immersive sound experiences. Most existing research on predictors of presence in immersive technology experiences has used quantitative (often correlation) experiments, in which presence levels are measured at different levels of the factor of interest. Qualitative studies (for example, interviews or focus groups), and mixed results studies (combining quantitative and qualitative data to enhance understanding) in this field are rare. A convergent mixed methods approach was used to investigate and answer the overarching research question. In this type of mixed methods design, two databases are created in parallel, one quantitative and one qualitative, before combining them for interpretation (Creswell & Plano Clark, 2018).

The real-life setting was another important aspect of the current study. Most relevant earlier studies have been conducted in a lab setting, making it important to test whether the findings will generalize to real-life situations. For example, the presence of ambient noise in real world settings may reduce any advantage of immersive sound.

**1.2 Research Questions**

The overall research question "Which variables predicted presence in this real-world immersive audio experience?" was addressed through a combination of more focused quantitative, qualitative and mixed methods research questions. Quantitative analysis examined the extent to which the factors of interest (technological immersion, audio quality, narrative, emotional and cognitive engagement, personal characteristics and environment) predicted presence. We hypothesised that headphones delivering spatialised sound would create higher presence than audio speakers delivering stereo sound. The strength of emotional and cognitive engagement, and previous knowledge were also expected to positively predict presence. As existing literature is mixed on a potential effect of age and gender on presence, no directional hypothesis was provided. Qualitative analysis focused on identifying factors that the interviewees felt helped to create their sense of presence. And the mixed methods analysis examined the extent to which the quantitative results, based on short Likert-scale measures, aligned with the qualitative findings.

**1.3 General Methods**

The object of interest in this study is the immersive sound installation 'One Story, Many Voices' (Figure 1), created for Imperial War Museums (IWM), as part of their Second World War and Holocaust Partnership Programme (SWWHPP, www.iwm.org.uk/partnerships/second-world-war-and-holocaust-partnership). Museum visitors could listen to eight different stories from different writers. The installation toured ten different locations in the UK from October 2021 until April 2023. The study was created working in close collaboration with the IWM / SWWHPP programme lead and the survey created with input from each of the partnership museums.

**Figure 1**

*The sound installation*

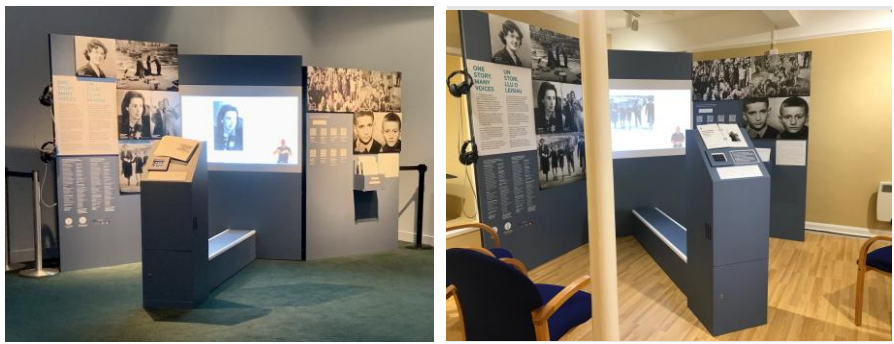

*Note*. This figure shows examples of the sound installation in situ. Image left taken at IWM London (December 2021), image right at Bodmin Keep (February 2022). The sound installation is the same, but the panel placement and seating arrangements were slightly different between locations.

Eight stories were created by professional creatives trained by the StoryFutures Writing Room to create binaural stories. Each story was created working with one of the museum partners (visit www.onestorymanyvoices.iwm.org.uk to listen to the stories). The stories were combined and played in a loop on the audio installation. Visitors could listen to the story that was playing when they entered the room or request a specific story via a number pad on the installation.

Participants could choose to listen to the stories through integrated headphones (MK I Armour Cable Headphones ACC-40, figure 2, left), or through a 2.1 stereo, speaker setup (Logitech Z333 speaker system with subwoofer) built into the base panel (figure 2, right). The headphones delivered spatialised sound, whereas the speakers delivered stereo sound.

**Figure 2**

Participants listening through headphones or speaker

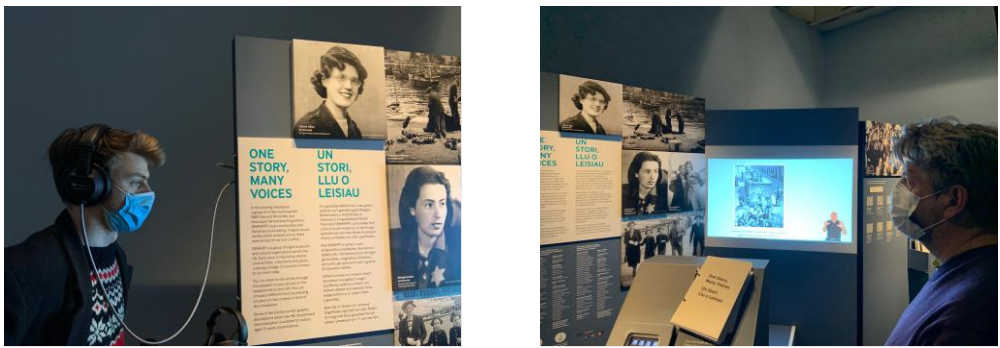

*Note*. The image shows a user listening to a story through one of the headphones (left) and integrated speaker (right)

The spatialised sound works were created by two sound designers. Each created four of the stories to be experienced through headphones. One used Ableton Live digital audio workstation (DAW) in conjunction with 'Dear Reality - dearVR PRO' audio plugin. This plugin can be used on individual audio tracks in the DAW for generic head-related-transfer-function (HRTF), binaural automation and rendering. It is capable of binauralisation of mono audio assets that can be automated around the predetermined listening position, across both the azimuth and elevation listening planes, resulting in a rendered linear binaural audio file that can be exported in stereo format. In order to represent a fuller, more three-dimensional (3D) listening experience, a combination of ambisonic and stereo sound beds plus reverb effects were implemented in separate audio tracks, alongside voice over and other story related audio assets, to create a rich and immersive listening experience. The other sound designer utilised the same immersive audio construction techniques as discussed above, however, used Logic PRO (DAW) and the in-built object 3D panner to achieve the generic HRTF binaural effect in the final output. These are similar technical audio creation pipelines, however, the algorithms used in the binaural/3D plugins are from different authors, which present marginal differences in spatial accuracy and delivery.

Participation was voluntary, and all participants gave informed consent before taking part. The study was run under the jurisdiction of Royal Holloway University of London Ethics Committee and in line with the Declaration of Helsinki. A risk assessment (including but not limited to Covid-19 safety) was completed before the study. Since the topics (Second World War and Holocaust) are highly sensitive, all locations had experienced volunteers on site if participants needed assistance. Sources of support were also provided at the end of the survey. As a thank you for survey participation, participants could choose to be included in a prize draw for a chance to win one of four £25 e-vouchers per location. Those who also participated in a follow-up interview were provided with a £15 e-voucher after interview completion. Those participating in the interviews provided additional informed consent before the interviews, which also included opt-ins for using their anonymised quotes for several purposes (for example, research papers and/or conferences).

## 2. Quantitative Research Strand

### 2.1 Quantitative Methods

All visitors to the museum who listened to at least one of the stories could participate in the research. Only those who used IWM speakers or IWM headphones were included in the analysis. People could also listen to the stories on location through their own phone, or anywhere on a website, but these participants were not included due to the technology being used being uncontrollably different from the headphones and speakers on site, and low sample sizes.

Data was collected for each of the ten locations separately, with online surveys filled out on a phone or iPad at the museum after listening to at least one of the stories, on the day of visit. Data collection was supported by museum volunteers (or the first author in two locations). The online survey was made accessible at the locations through a QR code and URL on a poster on the wall near the sound installation, and on flyers distributed on location to visitors. Both the QR and URL took users to the online survey, so that they could fill out the survey on their own smartphone. If participants could not use their own phone, an iPad with the survey pre-loaded could be used to fill out the survey. A volunteer briefing video and transcript were provided to the museums to ensure adequate and consistent training of volunteers. For an example of the briefing copy and other elements of the research support package provided to the museums, see Supplement A. An online training session with each of the museum leads and/or their volunteer leads was held before data collection started to discuss the research objectives, requirements, training materials and support package available, and to ask for up to three bespoke questions for their location's survey. Data was collected using the StoryFutures (part of Royal Holloway University of London) testing platform testxr.org.

After data cleaning 185 surveys were included. Survey participants ages ranged widely from 16 to 81 (see Supplement B), most participated in IWM London, Museum of Cornish Life, or Discovery Museum Newcastle (see Supplement B), 61% were female, 32% male (7% unknown), participants were generally educated (24% GCSE or A-levels, 39% Cert/diploma/BA, 21% MA/PhD, 16% unknown), most came from the UK (69% UK, 13% other, 17% unknown) and most were white (83% white, 16% unknown/other). Of the 185 surveys, 128 were from people using speakers and 57 from those using headphones. As participants were able to select their preferred audio device (speaker or headset), it was important to test if the two resulting self-selected groups were similar enough demographically for their data to be combined for analysis. Analyses (Independent-Samples Mann-Whitney U Tests for age and chi-square test of independence for the other variables) confirmed that the groups' demographics (for example, age, gender, education, country of residence) were highly similar.

The full survey is provided in Appendix A, the survey informed consent in Supplement C. In the survey, the dependent variable (DV) of 'presence' was measured using four questions taken from the Igroup Presence Questionnaire (IPQ, Schubert et al., 2001; Igroup, 1995–2016). The full IPQ questionnaire is validated and used in several immersive environments, including the audio study by Brinkman et al. (2015) mentioned earlier. For brevity one IPQ question per subscale (spatial presence, involvement, sense of being there and realism) was selected (see Verhulst et al., 2021 for a previous use of this approach). The four questions were further adapted for the audio-only nature of the experience by replacing the term 'virtual world' with 'audio world'. The resulting four questions were measured on five-point Likert scales: spatial presence ('I felt present in the audio world', with response options ranging from fully disagree to fully agree); involvement (with 'I was completely captivated by the audio world', from fully disagree to fully agree); sense of being there ('in the audio world I had a sense of "being there"', from fully disagree to fully agree), and realism ('how real did the audio world seem to you?', from completely real to completely unreal, in line with the original scale order from Hendrix (1994) used in the IPQ). Item score options were from 1 for fully disagree or completely unreal, to 5 for fully agree or completely real. Presence (average) was calculated as the unweighted average of the four presence questions and used as the DV.

Participants were able to listen to multiple stories as part of the installation (see Supplement B), with most listening to either one story (37% of participants), two (40%) or three stories (11%). If they listened to more than one story, the survey asked them to reply in relation to a single story. They were asked how much of the selected story they had listened to, and their data was excluded if they had not listened to at least 'most of' the chosen story.

In the survey, each potential predictive factor was measured with several questions (see Appendix A). Two audio related questions were taken from the Swedish Viewer-User Presence (SVUP-short) questionnaire (Larsson et al., 2001); one about clarity ('To what extent were you able to identify sounds?') and the other about contribution to realism ('To what extent did you think that the sound contributed to the overall realism'?). The emotion measurements were based on a condensed version of the 'discrete emotion questionnaire' (Harmon-Jones et al., 2016), with less relevant emotions, given the wartime focus of the narratives, (e.g., happy) replaced by relevant ones (e.g., sympathy/empathy). Cognitive engagement was measured with items (e.g., 'I was totally absorbed in what I was doing') taken from the flow short scale by Rheinberg et al. (2003), which has been validated in the context of computer games (Weibel & Wissmath, 2011). In addition to the seven factors under study, participants were also asked to rate their overall audio installation experience, in line with existing studies (Weibel & Wissmath, 2011). An attention check, similar in format to the other questions, was included halfway through the survey; those who failed the question (in other words, did not pay attention to the questions) were removed from the dataset. Survey duration was measured automatically; data of those participants who completed the survey in less than 3 minutes were removed. Because hearing loss may affect sound perception, we asked "Do you have hearing loss?" (Demeester et al., 2012); anyone reporting a significant level of uncorrected hearing loss was excluded from the dataset. Data from those who did not specify whether they had used headphones or speakers was also excluded. After the first location (IWM London) small survey optimizations were made for clarity, after which the survey remained the same with the addition of a small number of location-specific questions. None of these changes affected the measures reported.

**2.2. Quantitative Results**

*2.2.1 Variable Selection*

Given the sample size (n=185) and the rule of thumb (10 participants per variable, White, 2022) up to 18 variables could be included in the multiple regression analysis. Only items that correlated significantly with presence (see Supplement D) were considered. Four variables were removed due to too much missing data (audio quality rating, sound contributed to realism, I could identify sounds and I felt empathy/sympathy; 33-58% missing data), with all other variables having acceptable levels of missing data (between 0 and 8%). Including multiple variables that are highly correlated (e.g., those related to the same construct) can inflate standard errors and p-values, which may increase the likelihood of these variables appearing statistically non-significant in multiple regression analyses (Gelman et al., 2020). To address this issue, we selected a single representative variable for each group of correlated variables. Specifically, the emotions 'curious,' 'moved/touched,' 'sad,' and 'upset' were excluded because they correlated strongly (r = 0.4-0.7) with the overall emotional item 'the experience affected me emotionally,' which was retained. Similarly, ratings of the story being 'understandable,' 'meaningful,' 'interesting,' and 'personally relevant' correlated strongly with the overall story rating (r = 0.5-0.7), so only the overall story rating was included. Ratings of 'I find it important that the combination of stories reflects the whole of the UK' was removed as it correlated only moderately with presence (r=.18, p<.02) and correlated with 'I enjoyed listening to the stories being told by people with different accents' (r=.32, p<.001), which was included.

To understand if the nominal variables gender and/or immersion technology should also be included in the multiple regression, we tested if presence levels were significantly different between male and female participants, and headphone and speaker users, respectively. As the data was non-normally distributed, non-parametric tests were used (Field, 2013). Regarding gender, the median presence level was the same for men and women (both Mdn=3.75), therefore it was not included in the multiple regression. Regarding technological immersion, a Mann-Whitney U test suggested that average presence was higher for those listening through headphones (which provided spatialised sound; Mdn = 4.00) than through speakers (which provided 2D audio; Mdn = 3.75, U = 3795.00, n = 174, p =.03, 1-sided). This difference is likely to reflect differences in the audio experience provided, so these were investigated further (see Figure 3). Perceived spatialisation (measured with agreement with 'Sounds seemed to come from sources located around me') and perceived realism (measured with agreement with 'The sound contributed to the overall realism') were significantly higher for those using headphones than those using speakers. The other audio aspects were not significantly different between the groups, although the absolute values were consistently higher for those using headphones. In addition, a Mann-Whitney U test suggested that the overall experience was rated significantly higher by those using headphones than speakers, U = 4118.50, p <.01 (1-sided), n=177.

**Figure 3**

*Audio Perception (Mean Ranks) for Headphones and Speakers*

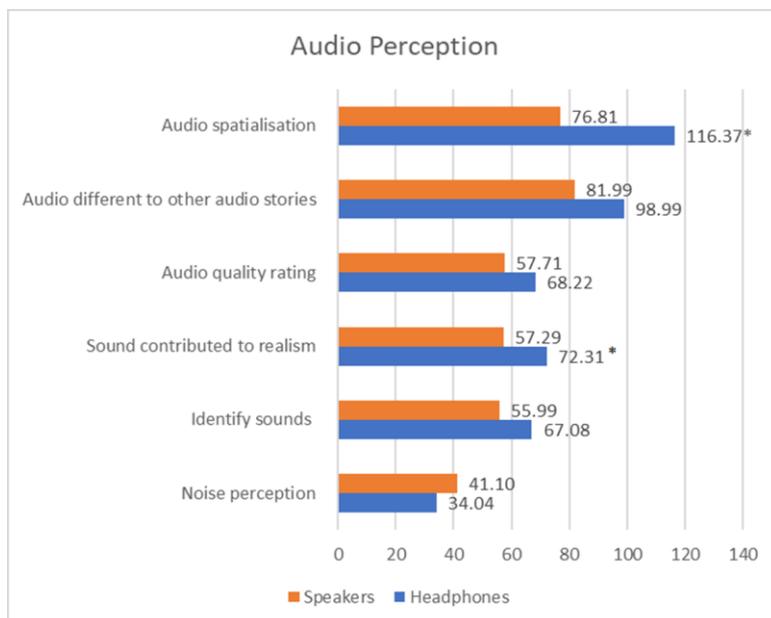

\* Denotes significance at the .05 level after Holm's (1979) correction for 6 tests.

Following the above, the following items were selected for inclusion in the analysis: presence (average of four presence items; dependent variable), and the predictor variables immersion technology (headphones or speakers), the overall installation rating (enjoyment), the audio items 'sounds seem to come from locations all around me', 'the audio was different to other audio experiences', 'I enjoyed listening to the stories being told by people with different accents', the cognitive engagement items 'I was totally absorbed', 'I wanted to know how the story ended', 'I felt a connection with the storyteller', and the story items 'overall story rating', 'the story had new content', 'the story was relevant to this location', and the emotional engagement item 'the story affected me emotionally'.

### 2.2.2 Multiple Regression

Multiple regressions were run to test which variables predicted presence. Since data was collected in different locations, it was possible that people within locations were and/or behaved differently from those in other locations. In other words, the statistical assumption that errors for participants are independent and identically distributed may not be tenable, as people within a location may be more like each other than to those of other cities. A Multilevel (also called mixed effects) Linear Model (MLM) with random intercept for location was tested, however this was not better than a classical multiple regression at predicting the data, according to the AICc (Akaike Information Criterion with correction for small sample sizes) and related BIC (Bayesian information criterion; AICs multimodal 225 vs classical 228, BIC both 268). Graphical inspection of model residuals did not reveal any obvious violations of normality and heteroscedasticity.

To understand which combination of variables created the best model (there were 4095 possible models given the number of variables), best model and variable importance analyses were run, with 'best' defined as the model with the lowest AICc, and variable importance was computed by summing the Akaike weights of all models that contained a given variable as predictor (Burnham & Anderson, 2002). The best model explained 53% of presence variability ($R^2$ = .53, $F(5, 125)$ = 28.12, $p < .001$, AIC = 223.16). As shown in Table 5, the variables that significantly predicted presence scores were: overall installation rating (enjoyment), the audio item 'the audio was different to other audio experiences', the cognitive engagement item 'I was totally absorbed' (indicating a state of 'flow'), the emotional / cognitive engagement item 'I felt a connection with the storyteller', and the story item 'overall story rating'. The other variables included in the multiple regression were of less importance when predicting presence, as shown in a variable importance analysis (see variables listed in order of importance in Table 6).

**Table 5**

*Best Model*

|  | b | SE | t | p |
|---|---|---|---|---|
| (Intercept) | 0.37 | 0.31 | 1.19 | .24 |
| Overall audio installation rating | 0.17 | 0.08 | 2.23 | .03* |
| The audio was different to other audio experiences | 0.16 | 0.06 | 2.72 | .01** |
| I was totally absorbed | 0.22 | 0.07 | 2.92 | <.01** |
| I felt a connection with the storyteller | 0.15 | 0.06 | 2.45 | .02* |
| Overall story rating | 0.19 | 0.09 | 2.11 | .04* |

Note. The table shows the variables that predict presence as part of the best (lowest AIC) model.
b = coefficient estimate, SE = Standard Error, t = t-statistic, p = probability.
***p< .001, **p<.01, *p<.05.

**Table 6**

*Variable Importance*

| Variable | Importance |
|---|---|
| I was totally absorbed | 0.98 |
| The audio was different to other audio experiences | 0.86 |
| Overall story rating | 0.82 |
| How would you rate the audio installation overall | 0.73 |
| I felt a connection with the storyteller | 0.69 |
| Sounds seemed to come from sources located around me | 0.56 |
| There was new content in the story | 0.44 |
| The story affected me emotionally | 0.43 |
| Technological immersion (headphones or speakers) | 0.39 |
| I wanted to learn how the story ended | 0.37 |
| Story relevance to the location I'm in now | 0.33 |
| I enjoyed listening to the stories being told by people with different accents | 0.26 |

*Note.* This table shows the relative importance of the predictors of presence, computed as the sum of Akaike weights across all models that include the specific predictor (Burnham & Anderson, 2002), with a higher value indicating a higher importance. It indicates how much the variable improves the model's fit to the data.

# 3. Qualitative Research Strand

## 3.1 Qualitative Methods

Semi-structured interviews were developed using the five phases from a systemic methodological review using ten high quality papers on creating semi-structured interviews (Kallio et al., 2016). The interview topics were in line with the survey to understand what influenced people's sense of presence, audio and story. We included questions that created the opportunity for new concepts to emerge (as per Kallio et al., 2016), such as, 'what else contributed to your sense of 'being there'?'. We also included questions about breaks in presence (BIPs), such as "Did your sense of 'being there' change throughout the experience?" and "Was there anything that interfered with or stopped your sense of being there?". The interview questions are included in Appendix B, the initial coding codebook in Supplement E and the interview guide in a Supplement F. Interviewees were self-selected from the survey participants; survey participants were asked to leave their email address to be contacted if they were interested in participating in a follow up online interview. Interview data was collected online via Zoom by the first author and recorded between March 2022 and February 2023. During the interviews, two participants declared working or having worked for a museum partner and were subsequently excluded, as they cannot be considered representative of the general audio installation visitor population. Therefore, nine of eleven interviews were included for analysis. The recorded audio files were transcribed using Trint (Trint Limited), manually checked and corrected where needed, and imported into NVivo 12 Plus (Lumivero). The interviews and interviewees' answers to open survey questions (for example, 'why did you give the audio rating') were added to the information for the relevant person from the interview transcripts, to be included in coding. Since the main objective of the study was to understand relationships between presence and predictor variables, relational analysis, a type of content analysis, was used (Busch et al., 1994-2023; Constable et al., 1994-2012). Interview participants' details are presented in Table 7, labelled by ID number for anonymity. Their demographic profile was similar to that of the survey participants.

**Table 7**

*Interview Participants*

| Number | Age | Gender | Education | Country | Ethnicity | Location | Audio Device |
|---|---|---|---|---|---|---|---|
| 1 | 16 | Female | Other | UK | White | Bodmin Keep Army Museum | Speakers |
| 2 | 64 | Female | BA | UK | White | IWM London | Speakers |
| 3 | 47 | Male | MA | UK | White | IWM London | Headphones |
| 4 | 19 | Male | BA | UK | white | Discovery Museum, Newcastle | Speakers |
| 5 | 56 | Male | BA | UK | Arabic | Ulster Museum, Belfast | Speakers on own phone |
| 6 | 18 | Male | A-levels | Ireland | White | Ulster Museum, Belfast & website at home | Own headphones & own phone |
| 7 | 32 | Female | PhD | Not reported | White | Ulster Museum, Belfast | Headphones |
| 8 | 42 | Female | GCSEs | UK | White | Manchester Jewish Museum | Speakers |
| 9 | 65 | Female | BA | UK | White | Manchester Jewish Museum | Speakers |

*Note*. This table shows details about each of the interview participants.

**3.2 Qualitative Results**

In relational analysis, relationships between concepts are coded (CSU, 2023). In the present study, this involved coding the relationships between presence and its predictors as explicitly discussed by interviewees. The predictive relationships between presence and its potential predictor variables were coded as 'variable x (for example, audio quality) was seen by a participant to influence their sense of presence', that is x —> presence, with the arrow indicating the participant perceived direction of influence, in other words, in this example x is seen to influence presence. Relationship coding was conducted primarily at the overall presence level. Where participants provided more elaborative responses, relations were coded as part of the qualitative analyses at the more detailed presence factor (for example, spatial presence), however this level of detail was not an objective in the qualitative part of the study. The amount of coding references in all relationships shows that the participants discussed many relationships with presence in their interviews and/or open survey answers. The most frequently referenced predictors are shown in Table 8. All relationships between the predictor concepts and presence discussed in the qualitative data are visualised in figure 4.

**Table 8**

*Most frequently mentioned presence predictors from qualitative data*

| Name of the predictor that interview participants stated as influencing their sense of presence. | How many times was this relationship mentioned? | In how many interviews and/or open survey answers was this relationship mentioned? |
|---|---|---|
| Nature of the audio experience | 28 | 15 |
| Visuals | 18 | 9 |
| Sound disturbances | 16 | 11 |
| Storyteller's tone of voice | 13 | 10 |
| Technological aspects (overall experience) | 13 | 7 |
| Authenticity (overall experience) | 13 | 6 |
| Soundscape (background sounds built into the audio experience) | 13 | 5 |

Note. This table shows the presence predictors most frequently mentioned in the qualitative data (interviews and open survey answers). The frequencies are totals of all participants combined. From these predictors, only 'sound disturbances' was mentioned by participants in relation to a specific presence factor (captivation), the other predictors were not mentioned in relation to a specific presence factor, in other words are seen in relation to presence (overall).

**Figure 4**

*Overview of Presence Relationships from Qualitative Data*

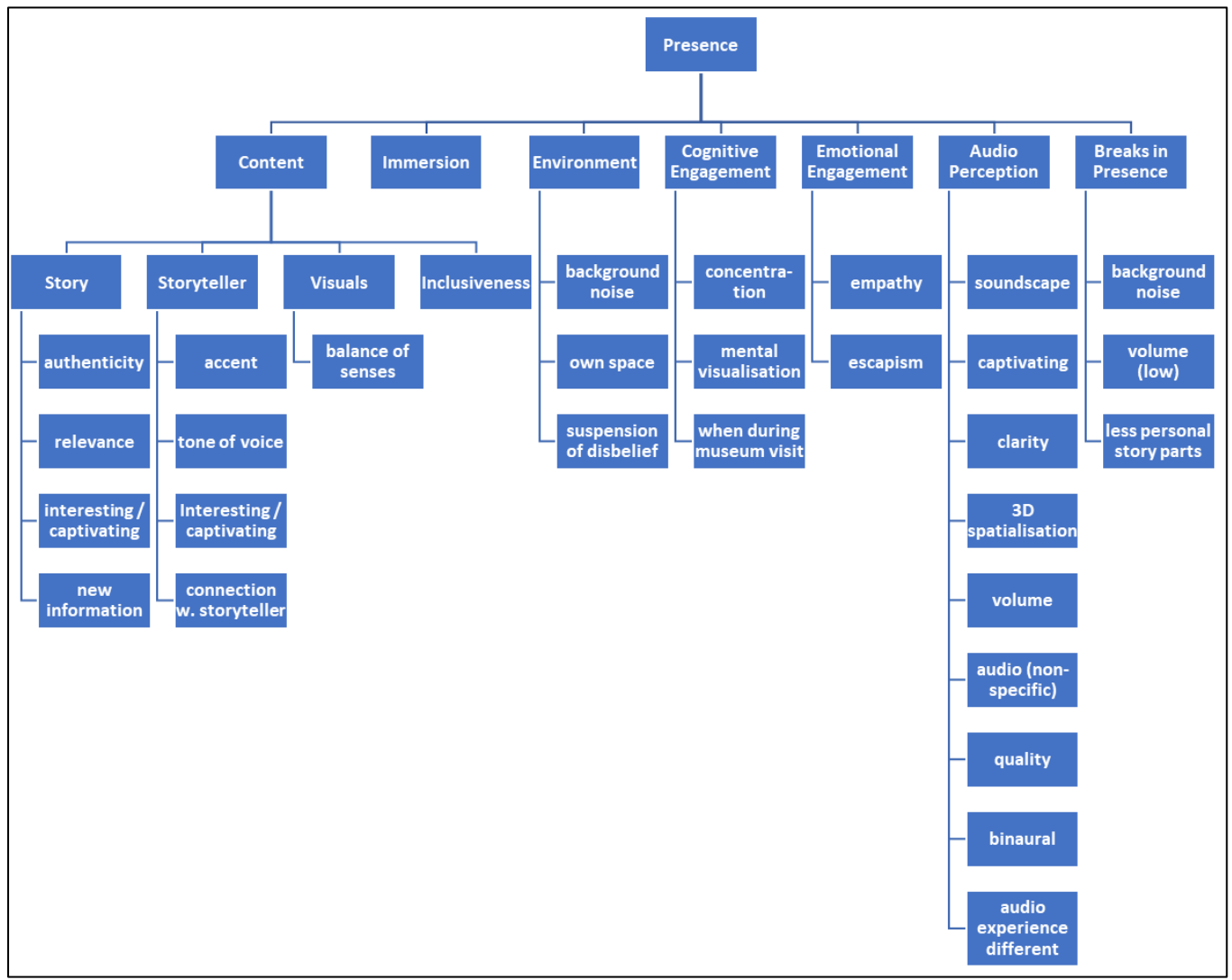

Note. This figure visualises the relationships between predictor concepts and presence that were explicitly mentioned by multiple interviewees.

The relationships between presence and each of its perceived predictors are exemplified below with interviewees' quotes (with additional quotes in Appendix C).

About the relationship between immersion technology (here, speakers or headphones) and presence, several participants stated that they felt high presence levels were partially due to using headphones (—> overall presence, 7 in 7). For example, P3 stated: "I don't think it would have been as powerful if you weren't on headphones. Because I think you just end up with the distraction… someone that's coming in, and there's bound to be a packet of crisps involved somewhere or something, you know, some other noise that just breaks you from that experience."

The positive relationship between the nature of the audio experience and presence was discussed most often at the overall presence level (audio —> presence overall, 11 in 5). For instance, "As soon as you got immersed in the audio, it was really powerful, you just really felt drawn into it. You were just totally part of the story" (P3).

Several narrative (content) aspects were perceived as influencing presence, specifically, the content of the stories, how they were told, visuals and inclusiveness. For example, the content of the stories included authenticity, relevance and new information. Authenticity was of paramount importance to the creation of presence (—> presence overall 5 in 4). For example, "The story wasn't just a story, it was someone's story, and it was real. I think was a real key factor [of being there]." (P3).

Emotional engagement was also seen to influence presence, specifically empathy (—> overall presence, 2 in 2), and escapism (overall presence, 3 in 2). For example, when talking about what created her sense of presence, P7 stated "Because he says he's just 15, […] I kind of empathise with that, how that would have felt. I think that helped."

Cognitive engagement was also mentioned as a predictor of presence (—> overall presence, 4 in 3), and more specifically, concentration (—> captivated presence factor, 2 in 2), mental visualisation (—> overall presence, 3 in 2) and the effect of when they listened during their visit (6 in 2). For example, some mentioned the effect of movies, for example, P9 saw mental images from the movie Dunkirk "I imagined in my head, [… it] puts pictures to the words that you're hearing, added more detail to [this story]."

The environment (locations) had a perceived impact on presence, both positively and negatively. In relation to positive impacts (—> presence overall, 5 in 3), P3 stated both the room the installation was in, and wearing headphones, helped to create his sense of presence: "You went in there, and you started that journey […] into the unknown. Then you put the headset on […] and you were immersed in it, you were there." On the other hand, sound disturbance negatively impacted some participants' presence in other locations (—> presence overall, 4 in 4). For example, P8 mentioned noise from reception "The background noise around the reception was a bit distracting. […] You're trying to get immersed and it's almost like someone pulling you into the screen, but you've got someone pulling you back a bit as well."

Personal characteristics were not discussed (nor spontaneously mentioned) in the qualitative data.

Following the qualitative analysis, the quantitative and qualitative findings were combined and compared to create mixed-methods findings (as per Creswell & Plano Clark, 2018). The findings are outlined in the discussion, and detail can be found in Appendix D. In addition, findings on Breaks in Presence (linked to the environment factor) and perceived differentiation of the presence survey questions (additional information from the interviews) can be found in Supplements G and H.

## 4. Discussion

This study found mixed methods support (quantitative from surveys and qualitative from interviews and open answer survey questions) to suggest that presence was predicted by audio quality (especially the audio being perceived as different from other audio stories participants had listened to in the past), cognitive engagement (especially being totally absorbed), the narrative experienced (the particular story and how it was told) and some support was found for emotional engagement (especially feeling a connection with the storyteller). The effect of the environment and immersion technology on people's sense of presence received only either quantitative or qualitative support. No support was found for any influence of personal characteristics.

**4.1. Mixed-Methods Findings**

We found in both the quantitative and qualitative data that headphones delivering spatialised sound created higher presence levels than audio speakers delivering stereo sound, which was as hypothesized. The higher overall presence levels for headphones may relate to the spatialisation of the sound that could be experienced using headphones, as suggested by significantly higher scores for headphone users than speaker users on both perceived spatialisation of sound ('sounds seemed to come from sources located around me') and the perception that the 'sound contributed to realism' (in line with its importance proposed by e.g., Magnenat-Thalmann et al., 2005). However, this finding may also relate to the general quality of the sound and/or overall experience being better for headphone users (in line with e.g., Pettey et al., 2008 who argued headphones may offer a more 'intimate' experience, and Bracken et al., 2010). The overall installation rating (enjoyment) was also significantly better when using headphones with spatialised sound than speakers with 2D audio, possibly influenced by the aforementioned differences in the experience afforded by headphones versus speakers. Additional research would be necessary to demonstrate an unequivocal contribution of spatialised sound to strengthening the sense of presence and/or improving overall ratings.

The links between visitors' audio experiences and their sense of presence were measured with survey questions and interviews, with a positive effect hypothesized. Mixed-methods support was found for the predictive relationship between audio experience and presence, especially for 'the audio was different, when compared to other audio stories I have listened to in the past'. In addition, several audio items positively correlated with presence (spatialisation, overall audio quality, 'sound contributed to realism', and 'I could identify the sounds'; as a reminder, the last 3 were not part of the quantitative analysis due to missing data). In the interviews further support was found for these items being linked to presence, plus the soundscape, the audio being perceived as captivating, the audio clarity and the volume.

The content (here, the nature of the narrative) was hypothesised to affect presence, in line with existing research (e.g., Felton & Jackson, 2022; Riches et al., 2019). Mixed-methods support was found for this, especially in relation to the overall story rating. In addition, several of the narrative-related items positively correlated with presence (namely, whether the story was understandable, interesting, meaningful, had new content, had personal relevance, and was relevant to the location, whether participants enjoyed the different accents and whether the stories reflected the whole of the UK). In the interviews, most of these items received further support. In addition, interviewees mentioned the importance for presence of the story being told with an authentic / passionate tone of voice, and the audio story being supported with visual elements (for example, images, BSL interpretation and subtitles); although there was some disagreement as to whether more visuals would have been helpful or distracting from the audio. The positive effect of the narrators on presence may be due to the voice talent selection being partially based on their match with the story, often in consultation with the partners and/or the writer. For example, the Cornish museums were keen to use a well-known Cornish voice to add to the sense of place.

Emotional engagement was hypothesised to positively predict presence. Mixed methods support was found for 'I felt a connection with the storyteller' which had both an emotional and a cognitive aspect. In addition, several emotional engagement items positively correlated with presence (the story affected me emotionally, I felt moved / touched, curious, empathy / sympathy, sad and upset). Although it was not a focus of the interviews to discuss the possible effect of emotions on presence, some participants did mention unprompted the effect of them feeling empathy and escapism on their sense of presence.

Cognitive engagement was hypothesised and found to positively predict presence. Mixed methods support was found in relation to being totally absorbed (a 'flow' item, which offers support for researchers who regard flow as a factor influencing presence) and feeling 'a connection with the storyteller' (which has both a cognitive/content and emotional aspect). Although not actively discussed in interviews, several participants mentioned that cognitive aspects assisted their sense of presence , including concentration, mental visualisation, the stage at which they listened to the story during their visit, and being able to let go of the knowledge of being physically present in a museum, to be virtually present in the audio story (i.e., suspension of disbelief, Slater & Usoh, 1993).

Personal characteristics (for example, age and gender) were not found to be predictors of presence in this study, adding another null finding to the inconclusive literature on this factor's possible effect on presence discussed in the introduction. This may be partially due to the museums' visitor samples not being diverse enough to capture differences in a wider population; although we had participants from a wide age range (16 - 81), over 90% were adults, and participants were generally educated, white and from the UK.

The nature of the environment was mentioned in the interviews as affecting presence; background noise was mentioned as affecting presence negatively in some locations, in line with Breaks in Presence (BIP) theory (Slater & Steed, 2000), whereas in other locations the installation having its own room was mentioned as having a positive effect on presence.

**4.2 Limitations**

It is important to acknowledge the limitations of this study. Firstly, because the audio installation was developed ahead of the research, it was not possible to separate the effect of spatialising the audio from the audio delivery method (speaker versus headphones), so it is not clear whether the user experience differences between the headphone and speaker conditions were created by the difference in spatialisation, the audio delivery tool (headphone vs. speaker), or a combination thereof. In future research, offering both stereo and immersive audio through headphones would isolate the audio type effect from the delivery method. For the same reason, randomisation of participants to conditions (headphones vs. speakers) was not possible in this setting, meaning that effects of self-selection (including the effect of Covid on their selection) may have played a part in the results. However, the demographics of those using headphones were very similar to those using speakers, so we were not able to identify any obvious confounds of this type. Lastly, on average, the interview participants reported higher presence than those who participated in the surveys only. Although we aimed to include interview participants with a similar range of presence scores, low interview signup during survey stage prevented this. Therefore, the interview outcomes may not be generalisable to all installation visitors. Importantly, in saying this, we did gain a depth of understanding from those who agreed to the interview about what created their sense of presence and how they re-established this following a break in presence. Nevertheless, it would also be valuable to interview participants with less sense of presence to understand their experience.

**4.3 Key Contributions of the Current Study**

This study addresses a gap in knowledge in the existing literature. For example, Felton and Jackson (2022) highlighted the need for additional research about location of audio sources, volume, and other auditory aspects that may predict presence. McRoberts (2018) suggested researchers pay attention to factors enhancing presence through the thematic narrative for a more complete understanding of presence. Although he mentioned this in relation to non-fiction VR experiences, we suggest a full understanding of the predictors of presence is similarly important in other types of immersive experiences, including immersive audio experiences. This study provides detailed support with high ecological validity due to the real-life setting, for the association and perceived influence on presence of several predictor variables that had been identified in the existing literature. It also examined a wide range of predictors, providing a more holistic approach than research which focuses on relationships between smaller sets of variables.

      Broadly speaking, we found that the factors that were found to predict presence in existing research (often lab-based VR studies) also held in this real-life, audio-only, setting. This is particularly interesting given that, at face value, the absence of immersive visuals in audio-only experiences would appear to constitute a profound difference in terms of the likely experience of 'presence' by comparison with VR. Nevertheless, the average presence levels reported were comparable with the levels that we have observed in a range of audiovisual VR experiences (average score for this audio experience: 3.8; average score across 6 VR experiences: 3.6). Given that visitors to heritage organisations are likely to want to see the sites and artefacts with their own eyes, the fact that a strong sense of immersive presence can be generated by high quality audio highlights this as a potentially impactful direction for future experiences.

# Appendix A Survey Questions

At time of submission, an example of the online survey from one of the locations (Devil's Porridge Museum) could be found at www.testxr.org/DPM, of which the items are listed below.

*Survey Overview*

**Survey Topic and Items**

**Audio**
Audio quality (overall rating)
Sound seemed to come from locations all around me
Sound contributed to realism
Could identify sounds
The audio was different compared to other audio stories

**Immersion Technology**
Headphones with spatialised sound / speakers with 2D sound

**Content**
Which story
Overall story rating
Story was understandable
Story was interesting
Story had personal relevance
Story was meaningful
Story was relevant to this location
Story had new content
Enjoyed different accents
Stories reflect whole UK

**Personal differences**
Age
Education level
Gender
Personal interests (for example, history and nature)

**Environment**
Noise perception

**Cognitive engagement**
I was totally absorbed in what I was doing
I felt a connection with the storyteller
I wanted to learn how the story ended
I knew what to do

**Emotional Engagement**
Affected me emotionally
Moved / touched
Curious
Empathy / sympathy

Relaxed

Sad

Upset

Fearful

Angry

Frustrated

**General Installation**

Overall installation rating (enjoyment)

Negatives (open answer box)

   *Note*. This table provides an overview of the survey's topics (in bold), and items per topic. This table was created for the journal article, not shown to participants.

   **Full Survey**

**Which story would you like to tell us about?** Please click on the corresponding image below. If you are not sure which image is for the story you want to tell us about, please see the booklet on the front of the installation, or on http://onestorymanyvoices.iwm.org.uk/stories.

How much of this particular story did you hear?

less than 1 minute

between 1 and 2 minutes

most of it

all of it

I listened to it more than once

Which, if any, of the stories did you listen to *in addition to* the one you have selected above? Please select all that apply.

How would you rate this **story**, using a 1-5 star rating, with 1 star being the lowest (worst) possible and 5 being the highest (best) possible rating?

★

★★

★★★

★★★★

★★★★★

Please briefly tell us why you gave this story rating.

How do you rate the story on each of these points? Please select your answer from the dropdown boxes. Do this once for every row.

very low / very bad

low / bad

neutral

high / good

very high / very good

COLUMNS: ['interesting', 'personal relevance', 'relevance to the location I'm in now', 'understandable', 'meaningful', 'there was new content in the story']

In the audio world I had a sense of "being there".

fully disagree

disagree somewhat

neutral

agree somewhat

fully agree

I felt present in the audio space.

fully disagree

disagree somewhat

neutral

agree somewhat

fully agree

I was completely captivated by the audio world.

fully disagree

disagree somewhat

neutral

agree somewhat

fully agree

How real did the audio world seem to you?

completely real

somewhat real

neither real nor unreal

somewhat unreal

completely unreal

I wanted to learn how the story ended

fully disagree

disagree somewhat

neutral

agree somewhat

fully agree

I was totally absorbed in what I was doing

fully disagree

disagree somewhat

neutral

agree somewhat

fully agree

The story made me aware of a different perspective and/or experience of war

fully disagree

disagree somewhat

neutral

agree somewhat

fully agree

The story affected me emotionally

fully disagree

disagree somewhat

neutral

agree somewhat

fully agree

This is an attention check question. Please select option 3 from the list below. Thank you.

I felt a connection with the individual telling the story

fully disagree

disagree somewhat

neutral

agree somewhat

fully agree

What did you think about the length of the story / stories you've listened to?

much too short

a little too short

about right

a little too long

much too long

To which extent did you feel each of the following emotions and feelings while listening to the story?

Rows: Curious, frustrated, moved/touched, sad, upset, sympathy/empathy

not at all

slightly

moderately

very much so

Any other emotions and/or feelings? (open)

To which extent do you agree with the following statement? "I find it important that the combination of stories reflects the whole of the UK."

fully disagree

disagree somewhat

neutral

agree somewhat

fully agree

How would you rate the **audio quality,** using a 1-5 star rating, with 1 star being the lowest (worst) possible and 5 being the highest (best) possible rating?

★

★★

★★★

★★★★

★★★★★

Please briefly tell us why you gave this audio quality rating?

Which of the audio tools below did you use to listen to the story you have told us about in this survey?

- the SPEAKERS built into the sound installation in the museum (no headphones)
- the HEADPHONES that are attached to the side of the sound installation in the museum
- my phone headphones
- my phone speakers

To which extent do you agree with the following statement? "Sounds seemed to come from sources located around me."

- fully disagree
- disagree somewhat
- neutral
- agree somewhat
- fully agree

Please briefly explain why you gave the rating on the previous question ('sounds seemed to come from sources located around me').

To which extent do you agree with the following statement? "The sound contributed to the overall realism"?

- fully disagree
- disagree somewhat
- neutral
- agree somewhat
- fully agree

To which extent do you agree with the following statement? "I was able to identify most or all of the sounds."

- fully disagree
- disagree somewhat
- neutral
- agree somewhat
- fully agree

Was the audio different, when compared to *other audio stories* you may have listened to in the past? THIS audio was...

- much worse
- a little worse

neutral (no different)

a little better

much better

Please briefly explain why you gave the rating on the previous question (if the audio was different to other audio stories).

To which extent do you agree with the following statement? "I enjoyed listening to the stories being told by people with different accents."

fully disagree

disagree somewhat

neutral

agree somewhat

fully agree

You may have heard some sounds from other people in the room around you. Please tick which of the below you agree with.

I heard sounds from other people and found them very disturbing

I heard sounds from other people and found them somewhat disturbing

I heard sounds from other people but did not find them disturbing

I did not hear sounds from other people

How would you rate **your experience with the overall audio installation**, using a 1-5 star rating, with 1 star being the lowest (worst) possible and 5 being the highest (best) possible rating?

★

★★

★★★

★★★★

★★★★★

Please briefly tell us why you gave this rating of your experience with the overall audio installation. (open)

To which extent do you agree with the following statement? "I reached new understanding or insight from my experience with the audio installation."

fully disagree

disagree somewhat

neutral

agree somewhat

fully agree

If you reached new understanding or insight, please can you briefly explain?

To which extent do you agree with the following statements?

fully disagree

disagree somewhat

neutral

agree somewhat

fully agree

COLUMNS: 'I will listen to more of these stories', 'This museum should have more audio experiences like this', 'I will tell my friends about this audio experience', 'This audio installation has made me more aware of the Imperial War Museum's partnerships with other organisations', 'The installation and the Scottish story within it made me feel recognised and proud', 'I learnt something new about the experiences of British Hondurans and/or evacuated orphans in the local area during the Second World War'.

To which extent has this audio experience changed your opinion of The Devil's Porridge Museum?

very negatively

negatively

neutral

positively

very positively

To which extent has this audio experience changed your opinion of Imperial War Museum (IWM) as a national museum of war and conflict?

very negatively

negatively

neutral

positively

very positively

Did you know what you were supposed to do and how things worked during this audio experience?

Did you experience any negative aspects, e.g. discomfort, technical difficulties, or other? Please type your answer in the box below. If you did not experience any negative effects please type none.

Age

Gender

What is your highest level of education?

GCSEs or equivalent

A-Levels or equivalent

Certificate or Diploma (L4-L5)

BA/BSc or equivalent

MA/MSc or equivalent

PhD or equivalent

Which country do you live in?

Please indicate which you agree with from the list below? Please tick all that apply.

I am interested in history and/or archaeology

I enjoy watching or listening to documentaries and factual programmes about history

I like to learn about military history

I enjoy visiting botanical gardens and/or nature reserves

I enjoy visiting flower / gardening shows

I like to understand about nature

I have a professional interest in the subject matter

Do you have hearing loss?

yes, corrected with e.g. a hearing aid

yes, NOT corrected with e.g. a hearing aid

don't know or prefer not to say

Would you say you are local to this museum?

yes

no

What is the first part of your postcode? Please provide the first letter(s) and number(s), e.g. DG12.

SENSITIVE QUESTION: It would help us to understand your ethnicity, but we understand this is a sensitive question, so please feel free to select 'prefer not to say' or skip this question if you prefer. The ethnicities are presented in alphabetical (neutral) order.

Asian / Asian British

Black / African / Caribbean / Black British

Mixed / multiple ethnic

SENSITIVE QUESTION: It would help us to understand if you are religious, but we understand this is a sensitive question, so please feel free to select 'prefer not to say' or skip this question if you prefer. The religions are presented in alphabetical (neutral) order.

Which of these have you visited *in the last 6 months* (including today)? Please select all that apply.

The Devil's Porridge Museum

The Museum of Cornish Life

Bodmin Keep Cornwall

Holocaust Survivors' Friendship Association at University of Huddersfield

IWM Imperial War Museum London

Aberystwyth University Arts Centre

Manchester Jewish Museum

National Holocaust Centre and Museum, Nottinghamshire

Tyne & Wear Archives & Museums

The website www.onestorymanyvoices.iwm.org.uk

Do you have any additional suggestions or comments?

Optional thank you prize draw: As a thank you for filling out this survey we would like to give you the opportunity to enter into a prize draw for one of four £25 Amazon vouchers. If you would like to enter into this prize draw please leave your email address in the box below. StoryFutures (on behalf of IWM) will only use this to inform the winners and send them their voucher, so you will not receive any spam or marketing information. Thank you.

Optional follow up interview: We are interested in asking 5-10 participants to further explain their opinion about the audio installation / stories in a follow up interview, which will be conducted online (for example via Zoom), after which an additional £15 thank you Amazon voucher will be provided. If you would like to be considered for participation in an online interview please leave your email address below. Please note this interview is optional, and it may be that not everyone who would like to can be invited. Thank you.

Thank you for your participation. We hope you have found participating in this study meaningful and interesting.

We are aware that these stories' topics may create strong emotions. If you feel upset and would like to talk about it, please contact your GP, or the Samaritans, a free 24/7 listening service. You can reach them on phone number 116 123 or https://www.samaritans.org/how-we-can-help/contact-samaritan/

If you have any questions about this research, please contact the researcher here. If you would like to be informed about future research opportunities by StoryFutures, please join their TestXR panel.

## Appendix B Interview Questions

- **Presence**:
    - PRESENCE LEVEL: Briefly review their presence answers from the survey;
        - I felt present in the audio world
        - I was completely captivated by the audio world
        - How real did the audio world seem to you
        - I had a sense of being there in the audio world
    - In summary, based on your answers of these survey questions, how would you describe to which extent you had a sense of being there in the audio world? Note to self: aim for rough level indication for example none / weak / moderate / strong sense of 'being there' to triangulate with the survey data and to check we cover a spread of presence levels across the interviews.
    - OPINION ABOUT THE PRESENCE QUESTIONS IN THE SURVEY:
      How did you interpret the four different questions we just discussed, while you were filling out the survey? In other words, did they seem to tap into different elements of your experience, or did they feel similar? … Why? Which? How?
- PREDICTORS:
    - What would you say were the most important aspects of the experience that created your feeling of 'being there in the audio world'?
      (Prompts if need be: For example, which of these aspects do you think were most important to create your feeling of 'being there'; the story, the narrator, the audio, the headphones / speakers, the installation, the museum setting, and/or anything else?)
    - What do you think would have improved your sense of being there in the audio world?
    - Anything else you'd like to tell me about this?
- BREAKS IN PRESENCE:
    - "Can you describe your sense of 'being there' throughout the experience (from the start to the end)"?
    - Did your sense of 'being there' change throughout the experience?
    - "Can you describe what aspects of your sense of being there changed or remained the same throughout the experience?"
    - Was there anything that interfered with or stopped your sense of being there?
      (Prompt if need be; for example, was there a moment in the experience that felt less real for you?)
    - If so:
        - What happened? When did this happen?
        - Why did this affect your feeling of being there?
        - Did the feeling of being there come back afterwards, or did it not come back? If it did come back – when / what happened then?
- Anything else you'd like to tell me about what may have affected your feeling of being there in the audio world?

- **Audio**
    - Review their survey audio data answers to these questions;
        - What did you think about the audio? – ask why
        - Perceived spatialisation of the sound – ask why
        - Was the audio better or worse when compared to other sound installations you may have listened to for example in museums? – if relevant, ask which / why
        - What could have made the audio better for you?
    - Anything else you'd like to tell me about the audio?

- **Story (if time allows)**
    - Discuss story rating / feedback – ask why they gave this rating?
    - DRIVERS: In rating the story, what elements did you factor in when making your judgements?
    - What could have made the story better for you?
    - Anything else you'd like to tell me about the story?

That was my last question. Can I check your email address, so I can send you your £15 Amazon voucher to thank you for your participation? …. Do you have any questions for me? *(Provide answer where possible.)*

**Debrief** (also provided on Testxr.org as electronic debrief sheet)

Thank you for your time and information, I greatly appreciate it. If you think of any additional questions about this research, please let me know on my email. This is also available on the webpage with the study information where you provided informed consent. Thank you, have a great day / evening. I will now end the interview and the recording. Goodbye.

*END & SAVE RECORDING & Arrange thank you voucher with IWM*

**Appendix C Participants Quotes per Presence Predictor**

The relationships between presence and each of its perceived predictors are exemplified below with interviewees' quotes. Although not all quotes explicitly mention this, all quotes are relevant to the relationship between the concept discussed and presence. Where interviewees used other words to describe their sense of presence (for example, 'immersion') their choice of words is maintained. Where there was disagreement between participants this is stated. Relationships discussed were mentioned by at least two participants. The seven predictors discussed in the introduction (immersion technology, audio, etc.) are discussed one at a time. Several of these relationships are interrelated (for example, volume, audio perception, background noise and cognitive engagement); however, this study's focus is on the relationships between the predictors and presence, rather than interrelationships between these predictors. Where arrows are used in the discussion, like before, they indicate the perceived direction of influence. Where two numbers are presented in brackets (for example, "x in y"), this means this relationship was mentioned x times, in y interviews or open survey answers (see similar use in Table 8).

**1 Perceived Relationship of Immersion Technology and Presence**

About the immersion technology (here, speakers or headphones), several participants stated that they felt high presence levels were partially due to using headphones (—> overall presence, 7 in 7). For example, P3 stated:

> "I don't think it would have been as powerful if you weren't on headphones. Because I think you just end up with the distraction... someone that's coming in, and there's bound to be a packet of crisps involved somewhere or something, you know, some other noise that just breaks you from that experience."

Further, P6 stated:

> "I think part of that [sense of being there] was the fact that I was wearing headphones at the time. I think it was the fact that I could hear certain sounds in one ear that I couldn't hear in the other. It was the surround sound sort of nature of it that added to that realism is brilliant."

Several of those using speakers stated they felt using headphones would have increased their presence, for example: "If I had [headphones with me to plug in], I'd have probably done that, because I think it would have gotten things a bit clearer" (P8, talking about aspects that helped to create that feeling of being there); "I saw some headphones, but I didn't know how to use them. So maybe if I knew how to use them would have helped" (P1, talking about what could further improve their sense of being there); "How it might feel to be completely submerged in the experience. I guess I could've used the headphones, but I didn't" (P9). However, P7 did not feel the headphone would increase presence:

> "For me it was more like because you've got the headphones on, you automatically are aware the sound's not coming from all around you, it's just coming from the headphones. Whereas if it was a walk-through experience with the sources of the sounds from different positions in an open room, you're a bit more immersed because you're just walking through."

**2 Perceived Relationship of Audio and Presence**

The positive relationship between the nature of the audio experience and presence was discussed most often at the overall presence level (audio —> presence overall, 11 in 5). For instance, "As soon as you got immersed in the audio, it was really powerful, you just really felt drawn into it. You were just totally part of the story" (P3) and "Obviously, the sounds, it enhances [the sense of being there]" (P8). More specifically, the soundscape affected the participants' presence positively (—> presence overall, 7 in 3, —> 'real' presence factor, 6 in 2, 'present' factor 2 in 1). When discussing the most important aspects that created that feeling of being there in the world, P7 stated "I think it would be the little bits of sound that you heard". For example, P6 mentioned:

> "Whenever she was getting out, throwing her leather satchel over her shoulder, you could hear the elastic stretching. That made the story come to life for me, it injected realism. […] I think the ambient sound could have been increased further"

Further, P8 explained how the soundscape impacted her sense of presence:

> "You could sort of almost see yourself at that time with all this going on around you, all the descriptions, the noises. I distinctly remember the like the click click of the horses' hooves and the air raid sirens. And you get goose bumps, I think, when you do hear that…"

The audio quality was positively linked with presence (—> present, 3 in 3), for example "I just really think it's the quality of the audio that lets you be immersed. You think 'I'm just in a different world'." (P3). The audio being captivating was also seen to increase presence (—> present factor, 2 in 2) for example, "the sound and the storytelling was really captivating, so I could still be in the story", as was the audio being clear (—> present factor, 2 in 2) for example "I was able to hear very clearly, which meant that I was able to be there" (P5).

The spatialisation of the audio was perceived to aid presence (—> presence overall, 4 in 2), for example, P6 explained:

> "Part of that [sense of being there] was the fact that I was wearing headphones at the time. … I could hear certain sounds in one ear that I couldn't hear in the other. It was the surround sound sort of nature of it that added to that realism. … Something that this audio did very, very well was that the sound was directly proportional to where it felt you were. For example, say a sound was going away from a person or sound was coming near… There were buses at one point, and that mirroring was very good. Compared to audio installations I've heard from another institutions, it was either directly where the headphones were or directly in front of me; there wasn't any spatial awareness, that didn't feel like I was in an environment audibly."

Whereas the previous predictors were generally seen as improving presence, the volume of the audio was mentioned frequently as being too low, which negatively impacted presence (—> overall presence, 4 in 3). For example, when asked about predictors of presence participants stated "The whole audio should be turned up a bit because you couldn't hear as well" (P1); "I had headphones at maximum volume, but there were bits where I sort of found myself drifting off because the volume was just not enough to push the sound to my front facing attention" (P6). However, not all participants felt the volume was too low; one participant stated, "the volume was enough to fill the room effectively" (P4). In this participant's location, the audio installation was placed in a quiet area of the museum, which was not possible in several other locations, so background noise may have played a role in the relationship between volume and presence.

**3 Perceived Relationship of Narrative (Content) and Presence**

Several narrative (content) aspects were perceived as influencing presence, specifically, the content of the stories themselves, how they were told, the visuals and their inclusiveness, each of which is discussed below.

One element of the experience content that participants mentioned was the content of the stories themselves, which included authenticity, relevance and new information. Authenticity was of paramount importance to the creation of presence (—> presence overall 5 in 4). For example, "The story wasn't just a story, it was someone's story, and it was real. I think was a real key factor [of being there]." (P3). Further, P9 explained:

"The first-hand stories [created being there] … It was the one on one's, the person saying, 'this is what it's like for me, this is how I experienced'. It wasn't just a travelogue, it was 'here's some fishermen, this is what they're doing, but actually, this is their story'. And that was compelling."

All but one participant spoke positively about the authenticity of the stories, however the one interview participant reporting low presence (P2) did not find the story authentic, for example, "as far as I knew, it wasn't his family history. It didn't resonate." The lack of authenticity was also linked with a perceived lack of personal relevance for her, as the daughter of a Holocaust survivor.

The story's relevance (personal relevance and/or relevance to the local area) was also perceived as a predictor of presence by other interviewees (personal relevance —> presence overall, 4 in 4, local relevance —> presence overall, 2 in 2, relevant today —> presence overall, 1 in 1). For example, "I remember when I was a little girl fishing with my grandmother, and I can imagine how it would have maybe been like." (P7).

New information in the story, including but not limited to different areas of the UK, was also seen as predicting presence (—> presence overall, 4 in 2), for example, "They [the stories] were short and sweet, but there was enough information in there, and the more you learnt, the more you're like, 'Wow, really, what?'" (P8) and "The stories were very different, and they were all somewhat unusual, it went below the usual history lesson. […] [What] was really captivating as well [were] the different characters and the fact that they were from different parts of the UK." (P9).

Another element of the experience's content that participants mentioned was the way in which the storyteller told the story. This was seen as affecting presence, especially the storytellers' accents (—> presence overall, 5 in 4), tone of voice (—> overall presence, 4 in 3), it being interesting / captivating (—> captivated, presence, 2 in 2), and feeling a connection with the storyteller (—> presence overall, 4 in 2). Regarding accents, P4 explained the importance for presence:

"Having the representative of more local dialects and voices definitely helped [to create that feeling of being there]. Having been able to hear something that would be more normal to your day-to-day hearing, it makes it feel more true and natural occurring, and it doesn't feel like you've been plopped right down in some random place and random time and listen to a story."

The storytellers' tone of voice was seen to affect presence, for example, "the precision of the words, and the tone of the voice, you were just carried on what must have been an absolutely harrowing journey" (P3); "The passion in the voice came across, you could buy into what they were saying." (P8). The stories and how they were told were perceived as captivating, which helped with presence. For example, when talking about what drove her sense of presence, P8 said "There was a big sort of pull just in terms of the story content, it was just pretty fascinating." Feeling a connection with the storyteller was also seen as important for presence, for example P4 said: "Being able to have a connection almost with the speaker, being able to relate to them, and being able to understand what they're saying, definitely helps with immersion."

Yet another element of the experience content that participants mentioned were the visual aspects, which were also seen to affect presence (—> overall presence, 6 in 5). For example, "the images that were displayed as well, you just really felt drawn into it" (P3). When discussing what created their sense of presence, P4 elaborated:

"Obviously, I was more engrossed in the audio aspect of it, but […] having visual aids that give you more of an idea of the time and what the speaker would have been seeing as well definitely helped".

Getting the sensory balance between audio and visuals right was seen as important for presence (—> overall presence, 6 in 4). Although some interview participants stated they would have liked more visuals, others stated the mix of mostly audio and a few supporting visuals was just right for them. On the one hand, P8 asked for more visuals, but also provided caution about the multisensory mix:

"I think people get more pulled in by things they see […] rather than things they hear. Something that grabbed me at the beginning was the poster, so if there were more visual things on the screen, it does kind of pull you in. But I think you've just got to be careful when you do that. The stories themselves, a lot of sympathy is generated for the people that are there. So, if you start putting too many images, too many graphics, it is kind of overloading and detracting from seriousness of it".

On the other hand, P3 mentioned two multisensory experiences that had negatively affected his sense of presence in the past:

> "It wasn't one of these, 'go and fly with the Red Arrows' type simulators. That's got smoke, it's got smell, it's got sound, it's got the vibrating floor and everything. In a way, you're not there. […] The Jorvik Viking Centre, I think, has got smell and stuff, but the smells didn't stay in the second place, so you just lose it. […] I just don't think this [audio experience] could have been told better."

The last element of the experience content that participants mentioned was the inclusiveness of the content. Offering subtitles and BSL interpretation was perceived as supporting presence, for example "For some reason, being able to read and being able to hear definitely made being able to hear it a lot better and helped with the immersion" (P4).

**4 Perceived Relationship of Emotional Engagement and Presence**

Emotional engagement was also seen to influence presence, specifically empathy (—> overall presence, 2 in 2), and escapism (overall presence, 3 in 2). For example, when talking about what created her sense of presence, P7 stated "Because he says he's just 15, […] I kind of empathise with that, how that would have felt. I think that helped." Escapism was mentioned by P6 when talking about what created his sense of presence:

> "It was the ability to take you from the real world, the sort of escapism. I escaped through her suffering, which is a weird way to put it, but it felt like I was listening along with [her]."

**5 Perceived Relationship of Cognitive Engagement and Presence**

Cognitive engagement was also mentioned as a predictor of presence (—> overall presence, 4 in 3), and more specifically, concentration (—> captivated presence factor, 2 in 2), mental visualisation (—> overall presence, 3 in 2) and the effect of when they listened during their visit (6 in 2). Regarding concentration, participants mentioned how sound and other environmental disturbances negatively affected their cognitive engagement and therefore their sense of being there. For example, P9 said:

> "Maybe because it was in the entrance hall of the Museum […] you can hear the cafe and [other] things going on. And [having] a couple of chairs so that you could have sat and really focussed because, [I worried] 'is someone going to trip over my bag, where I put my coat'. […] That's why 'being there' was rated less high, only for those sorts of environmental disturbances which interfered with my focus and concentration."

Several participants mentioned seeing mental pictures in their head while feeling present, for example, P8 explained how the way the story was told helped to create mental images that supported feeling present in the story:

> "It was the way that it was delivered, the sort of picture was painted; 'this is the scene, this is kind of you here, this is what's going on sort of around you'. And you did conjure up these mental images of it as it was going on."

In addition, some specifically mentioned the effect of movies, for example, P9 saw mental images from the movie Dunkirk "I imagined in my head, [… it] puts pictures to the words that you're hearing, added more detail to [this story]." and P7: "That's quite similar to that movie that I watched. […] I felt more strongly connected to it because of things I've seen in other types of media."

The point in time during their visit to the museum at which participants listened to the stories seems to have affected whether they were in the right mindset to fully engage with the experience, in turn affecting their sense of presence. For example, P8 stated "It was better to have watched it at the end [of our visit] as I think you get more out of it. […] When I've just walked in, I'm not necessarily in that right kind of mindset. So, take a deep breath and switch off from everything else, sort of be drawn into it." However, P2 stated "We had just been to the Holocaust Exhibition, […] seen all these really personal things. Then we just popped in [to the audio installation] before we were leaving, […] and I just felt 'I'm not sure what this is all about'."

**6 Perceived Relationship of the Environment and Presence**

The environment (locations) had a perceived impact on presence, both positively and negatively. In relation to positive impacts (—> presence overall, 5 in 3), P3 stated both the room the installation was in, and wearing headphones, helped to create his sense of presence:

> "You went in there, and you started that journey […] into the unknown. Then you put the headset on […] and you were immersed in it, you were there."

On the other hand, sound disturbance negatively impacted some participants' presence in other locations (—> presence overall, 4 in 4). For example, P8 mentioned noise from reception "The background noise around the reception was a bit distracting. […] You're trying to get immersed and it's almost like someone pulling you into the screen, but you've got someone pulling you back a bit as well." And P4 mentioned "hearing kids running around shouting, it definitely would take [you] out of the immersion".

Several participants mentioned they struggled to feel present in the story while they were physically present in a modern-day setting (—> spatial presence, 3 in 3). This struggle is known in the literature as suspension of disbelief (Slater & Usoh, 1993). For example, P5 said "I had to work hard to stop the museum location interfering with me feeling that audio experience was real. I couldn't be 100% that the story was real because I couldn't be in the museum and somewhere else." Similarly, P4 stated "I didn't feel as immersed because being in a museum with a bunch of modern people around. […] It's incredibly hard to immerse someone in a different era whilst they're on their mobile phone.".

**7 Perceived Relationship of the Environment and Presence**

Personal characteristics were not discussed (nor spontaneously mentioned) in the qualitative data.

**Appendix D Comparison of Quantitative and Qualitative Findings**

| Possible Predictors of Presence | Quantitative Analyses Findings | | | Qualitative Analysis Findings | Mixed Methods (integrated) Findings |
|---|---|---|---|---|---|
| | Presence significantly different between *categories*? | *Correlated* significantly with presence? | *Predicted* presence significantly in multiple regression? | Mentioned by more than one participant as a predictor of presence? | Is the predictive relationship between this variable and presence supported by both quant. and qual. findings? |
| **Audio** | | | | | Yes, especially on 'audio was different compared to other audio stories' |
| Audio quality (rating) | | YES | a | YES | |
| Sound seemed to come from locations all around me | | YES | NO | YES | |
| Sound contributed to realism | | YES | a | YES | |
| Could identify sounds | | YES | a | YES (as clarity and volume) | |
| Audio was different | | YES | YES | YES | |
| Binaural aspect | | | | YES | |
| Captivating audio | | | | YES | |
| Soundscape | | | | YES | |
| Clarity | | | | YES | |
| Volume | | | | YES | |
| **Immersion Technology** | | | | | Only qualitative support |
| Headphones (spatialised sound) or speakers (2D sound) | YES | | NO | YES | |

| Content (narrative) | | | | | Yes, although from different quant and qual items. |
|---|---|---|---|---|---|
| Overall story rating | | YES | YES | | |
| Story was understandable | | YES | b | | |
| Story was interesting | | YES | b | YES (as interesting, captivating) | |
| Story had personal relevance | | YES | b | YES | |
| Story was meaningful | | YES | b | YES (as 'story is authentic') | |
| Story was relevant to this location | | YES | NO | YES | |
| Story had new content | | YES | NO | YES | |
| Enjoy different accents | | YES | NO | YES | |
| Stories reflect the whole UK | | YES | b | YES (as 'different UK areas') | |
| Tone of voice | | | | YES | |
| Visuals | | | | YES | |
| Sensory Balance | | | | YES | |
| Inclusiveness | | | | YES | |
| **Personal differences** | | | | | No support found |
| Age | | NO | | | |
| Education level | | NO | | | |
| Gender | NO | | | | |
| **Environment** | | | | | Support only from qualitative results |
| Noise perception | | NO | | YES | |
| Own space for installation | | | | YES | |

| Cognitive engagement | | | | | Yes, especially on 'I was totally absorbed' |
|---|---|---|---|---|---|
| I was totally absorbed | | YES | YES | YES | |
| I wanted to learn how the story ended | | YES | NO | Implicitly | |
| I knew what to do | | | | Implicitly | |
| Concentration | | | | YES | |
| Mental visualisation | | | | YES | |
| When during museum-visit | | | | YES | |
| Suspension of disbelief | | | | YES | |
| **Emotional Engagement** | | | | | Yes, especially on 'I felt a connection with the storyteller' |
| I felt a connection with storyteller | | YES | YES | YES | |
| Affected me emotionally | | YES | NO | | |
| Moved / touched | | YES | b | | |
| Curious | | YES | b | | |
| Empathy / sympathy | | YES | a | YES | |
| Relaxed | | NO | | | |
| Sad | | YES | b | | |
| Upset | | YES | b | | |
| Fearful | | NO | | | |
| Angry | | NO | | | |
| Frustrated | | NO | | | |
| Escapism | | | | YES | |
| **Overall Installation** | | | | | Support from quant., not discussed in qual. |
| Overall audio installation rating | | YES | YES | | |
| Number of negatives mentioned | | NO | | | |

| **Breaks in Presence** | | | Support from qual, not measured in quant. |
|---|---|---|---|
| Noise perception (background noise) | See 'environment' above | YES | |
| Volume | | YES | |
| Less personal story parts | | YES | |

*Note*. This table shows the quantitative (surveys) findings and qualitative (interviews & open survey answers) findings per predictor of presence. On the right, they are compared, to understand which predictors are supported by both quantitative and qualitative data, in other words, the combined, or mixed-methods findings. Greyed out cells are deemed 'not applicable', for example, as not relevant for or not included in the type of quantitative analysis, or not a topic of discussion in the interviews. [a] This variable was not included in the multiple regression due to a small number of participants filling out this question. [b] This variable was not included in the multiple regression due to a strong correlation with another variable that was included.